\documentclass[english,aps,preprint]{revtex4}
\usepackage[]{fontenc}
\usepackage{amssymb}

\makeatletter

\usepackage{babel}
\makeatother
\begin{document}

\preprint{BROWN-HET-1383}

\preprint{ICN-UNAM-03/15}

\preprint{NSF-KITP-03-107}

\title{A new twist on dS/CFT}

\author{Alberto G\"uijosa}

\email{alberto@nuclecu.unam.mx}

\affiliation{Departamento de F\'{\i}sica de Altas Energ\'{\i}as, Instituto de
Ciencias Nucleares, Universidad Nacional Aut\'onoma de M\'exico,
Apdo. Postal 70-543, M\'exico, D.F. 04510, M\'exico}

\author{David A. Lowe}

\email{lowe@brown.edu}

\affiliation{Physics Department, Brown University, Providence, RI 02912, USA}

\begin{abstract}
We stress that the dS/CFT correspondence should be formulated using
unitary principal series representations of the de Sitter isometry
group/conformal group, rather than highest-weight representations
as originally proposed. These representations, however, are infinite-dimensional,
and so do not account for the finite gravitational entropy of de Sitter
space in a natural way. We then propose to replace the classical isometry
group by a $q$-deformed version. This is carried out in detail for
two-dimensional de Sitter and we find that the unitary principal series
representations deform to finite-dimensional unitary representations
of the quantum group. We believe this provides a promising microscopic
framework to account for the Bekenstein-Hawking entropy of de Sitter
space.
\end{abstract}
\maketitle

\section{Introduction}

A holographic formulation of quantum gravity in de Sitter space has
been proposed in \cite{Strominger:2001pn} (and anticipated in \cite{Park:1998yw,Park:1998qk}),
and the details of this correspondence have been elaborated further
in \cite{Bousso:2001mw,Spradlin:2001nb,Strominger:2001gp,Ghezelbash:2001vs,Li:2001pj,Balasubramanian:2001nb,Klemm:2001ea,Das:2002he,Kabat:2002hj,Leblond:2002tf,Larsen:2002et,Balasubramanian:2002zh,Kristjansson:2002yb,Goldstein:2003qf,Larsen:2003pf,Nojiri:2001mf}.
The basic picture conjectures that quantum gravity in de Sitter is
holographically dual to a boundary conformal field theory, that can
be viewed as residing at the spacelike past (and/or future) infinity.
The isometry group of de Sitter space is identified with the conformal
group on the boundary.

In another set of developments, holographic bounds on the entropy
in de Sitter space have been formulated, with the conclusion that
$S\leq A/4$, where $A$ is the area of the cosmological horizon \cite{Fischler:1998st,Easther:1999gk,Bak:1999hd,Banks:2002wr,Banks:2003ta,Bousso:1999cb,Bousso:1999xy,Bousso:2000nf,Goheer:2002vf,Kaloper:1999tt,Veneziano:1999ts}.
For eternal de Sitter, one can view this relation as bounding the
number of states accessible to a local observer. Exactly what one
means by accessible is then open to debate. Banks \emph{et al.} \cite{Bousso:2000nf,Banks:2002wr,Banks:2003ta}
have argued for the strongest interpretation of the bound, where the
number of states in the Hilbert space is $e^{A/4}$. Susskind \emph{et
al.} \cite{Goheer:2002vf,Dyson:2002nt} have instead argued the number
of states should be countably infinite, with the number of states
below a certain energy being bounded by $e^{A/4}$. Another possible
interpretation compatible with semiclassical physics is that the spectrum
of states is continuous, and that the gravitational entropy $A/4$
is simply a finite contribution to an otherwise infinite number of
accessible states. 

In the first two scenarios, there is apparent conflict with the dS/CFT
proposal. The isometry group of de Sitter involves non-compact boost
generators, yielding unitary irreducible representations that are
infinite dimensional. This conflicts with finiteness of the entropy.
A related argument given in \cite{Goheer:2002vf} studies a matrix
element of a boost $e^{iL(t)}$ assuming finiteness of the entropy
and hermiticity of the Hamiltonian of a comoving observer, and shows
this must be quasi-periodic as time $t$ is taken to infinity. The
symmetry group instead implies the matrix element has a finite constant
limit. One must then give up either hermiticity and/or the classical
symmetries to obtain a finite entropy. In this paper we take the view
that the unitarity of the description is paramount, and are ultimately
led to construct a formalism where the classical symmetries are relinquished. 

Our analysis begins at the classical level, with a review in Section
\ref{sec:review} of the original dS/CFT correspondence. In Section
\ref{sec:newdS/CFT} we then emphasize the point made in \cite{Balasubramanian:2002zh}
that the quantization of a scalar field on de Sitter space with mass
larger than the Hubble scale yields a unitary principal series representation
of the classical isometry group, as opposed to the standard non-unitary
highest-weight representations considered in the original dS/CFT conjecture.
It is then natural to propose a new version of the correspondence
using the principal series representations of the conformal group. 

However, principal series representations are infinite-dimensional,
conflicting with finite de Sitter entropy. We therefore propose to
replace the classical isometry group by a quantum symmetry group.
(For an introduction to quantum groups, see \cite{Klimyk:1997eb,Chari:1994pz}.)
The quantum group involves a deformation parameter $q$, which we
take to be a root of unity. The classical symmetry group is recovered
in the limit $q\to1$. Quantum group deformations have appeared in
the context of AdS/CFT in \cite{Jevicki:1998rr,Ho:1999bn,Steinacker:1999xu}.

For simplicity we will restrict our attention to the case of two-dimensional
de Sitter space; however, we expect the main results to generalize
to other dimensions. In Section \ref{sec:qdscft} we identify finite-dimensional
irreducible representations of the quantum group that become the unitary
principal series representations of the conformal group in the classical
limit. We propose a duality between a $q$-deformed conformal field
theory and a bulk gravitational theory in a $q$-deformed de Sitter
background. Bulk fields corresponding to these representations are
then to be identified with corresponding operators in the $q$-CFT.
This new formulation has many of the ingredients needed to resolve
the entropy finiteness issue mentioned above, and we discuss the analog
of the quasi-periodic correlator calculation of \cite{Goheer:2002vf}
in this new framework. In section \ref{sec:Discussion}, we end with
some discussion of topics such as the $q$-deformed CFT at the interacting
level, geometry of $q$-deformed de Sitter, the formulation of gravity
in this background, entropy calculations, and the higher dimensional
case. Finally, we warmly recommend \cite{Pouliot:2003vt,Jevicki:1998rr}
where many related ideas appear.

\section{Review of the Classical ds/cft Correspondence \label{sec:review}}

The classical dS/CFT correspondence was proposed by Strominger \cite{Strominger:2001pn}
and studied further in \cite{Bousso:2001mw}, whose notation we follow.
Thus far the correspondence does not go as far as specifying the CFT
at the interacting level. In the current state of development the
correspondence specifies a mapping from free bulk scalar field correlation
functions to boundary correlation functions of conformal primary operators. 

In terms of global coordinates, the metric on $d$-dimensional de
Sitter space is\[
\frac{ds^{2}}{L^{2}}=-dt^{2}+\cosh^{2}t~d\Omega_{d-1}^{2}~,\]
with $L$ the Hubble length, which will be henceforth set equal to
unity. 

The correspondence of \cite{Bousso:2001mw} was formulated for a scalar
field of mass $m$ in a general $\alpha$-vacuum. However, as described
in \cite{Goldstein:2003qf}, only the Euclidean vacuum state ($\alpha=-\infty$)
can be consistently coupled to gravity, so we will restrict our attention
to that case. On $\mathcal{I}^{-}$ and $\mathcal{I}^{+}$ the field
behaves as\begin{eqnarray}
\lim_{\tau\to-\infty}\phi(t,\Omega) & = & \phi_{+}^{in}(\Omega)e^{h_{+}t}+\phi_{-}^{in}(\Omega)e^{h_{-}t}\nonumber \\
\lim_{\tau\to\infty}\phi(t,\Omega_{A}) & = & \phi_{+}^{out}(\Omega)e^{-h_{+}t}+\phi_{-}^{out}(\Omega)e^{-h_{-}t}\label{eq:asyfield}\end{eqnarray}
where \begin{equation}
h_{\pm}=\frac{d-1}{2}\pm i\mu,\quad\mu=\sqrt{{m^{2}-\frac{(d-1)^{2}}{4}}}\label{eq:condim}\end{equation}
 and $\Omega_{A}$ denotes the point antipodal to $\Omega$ on the
$d-1$ sphere.

The correspondence proceeds by examining the bulk Wightman function
in the limit that points are taken to infinity\begin{eqnarray}
\lim_{\tau,\tau'\to-\infty}G_{E}(x,x') & = & e^{h_{+}(t+t')}\Delta_{+}(\Omega,\Omega')\label{eq:btwo}\\
 &  & +e^{h_{-}(t+t')}\Delta_{-}(\Omega,\Omega')~,\nonumber \end{eqnarray}
where $\Delta_{\pm}$ is proportional to the two-point function on
the sphere of a conformal primary of weight $h_{\pm}$. The expressions
(\ref{eq:btwo}) determine two-point functions of operators in the
dual CFT\begin{equation}
\left\langle 0|\mathcal{O}_{\pm}(\Omega)\mathcal{O}_{\pm}(\Omega')|0\right\rangle =-\frac{\mu^{2}}{4}\Delta_{\pm}(\Omega,\Omega')\label{eq:cfttwo}\end{equation}
together with the contact terms\begin{eqnarray}
\left\langle 0|\mathcal{O}_{-}(\Omega)\mathcal{O}_{+}(\Omega')|0\right\rangle  & = & \frac{\mu}{4}\delta^{(2)}(\Omega,\Omega')\label{eq:contact}\\
\left\langle 0|\mathcal{O}_{+}(\Omega)\mathcal{O}_{-}(\Omega')|0\right\rangle  & = & -\frac{\mu}{4}\delta^{(2)}(\Omega,\Omega')~.\nonumber \end{eqnarray}
We note that for bulk fields with masses $m>(d-1)/2$ (in units of
the Hubble scale), which we will concentrate on here, the quantity
$\mu$ is real and so the corresponding conformal primaries have complex
conformal weights. 

For the case of $dS_{3}$, the two-point functions in the CFT (\ref{eq:cfttwo})
can be used to define the Zamolodchikov norm in the standard way\begin{equation}
\left\langle \mathcal{O}_{\pm}(\Omega)|\mathcal{O}_{\pm}(\Omega)\right\rangle =\Delta_{\pm}(\Omega,-\Omega_{A})~.\label{eq:zamnorm}\end{equation}
 This norm corresponds to an inner product defined on fields in the
bulk that involves an additional action of $CPT$ as compared to the
standard Klein-Gordon inner product \cite{Witten:2001kn,Bousso:2001mw}. 

For future reference let us define precisely what we mean by unitarity
in this context \cite{Steinacker:1998kv}. A hermitian inner product
satisfies $\langle c\psi|\chi\rangle=\bar{c}\langle\psi|\chi\rangle=\langle\psi|\bar{c}\chi\rangle$
for $c\in\mathbb{C}$, together with the condition $\overline{\langle\psi|\chi\rangle}=\langle\chi|\psi\rangle$.
This inner product is said to be invariant if $\langle\psi|g\chi\rangle=\langle g^{*}\psi|\chi\rangle$
where $g$ is an element of the algebra, and $g^{*}$ is the adjoint
of $g$. Unitarity adds the condition that the hermitian inner product
be positive definite.

Representations with complex conformal weights are therefore non-unitary
\cite{Ginsparg:1988ui} with respect to the inner product (\ref{eq:zamnorm}),
for which $L_{n}^{*}=L_{-n}$. For example\[
\left\langle L_{0}h_{+}|L_{0}h_{+}\right\rangle =\left\langle h_{+}|L_{0}^{2}|h_{+}\right\rangle =h_{+}^{2}\langle h_{+}|h_{+}\rangle\]
which is not positive definite for nontrivial $|h_{+}\rangle.$

\section{New classical dS/CFT Proposal \label{sec:newdS/CFT}}

As we have seen, the conformal field theory as defined above is not
unitary. This issue arose because one insisted on establishing a correspondence
with the standard highest-weight representations considered in conformal
field theory. We propose instead to formulate the correspondence using
the unitary principal series representations of the de Sitter isometry
group/conformal group $SO(d,1)$. Closely related proposals appear
in \cite{Balasubramanian:2002zh}, though we differ on some of the
details and interpretation. 

To keep the discussion as explicit as possible, we will now restrict
our attention to two-dimensional de Sitter space. The classical de
Sitter isometry group is $SO(2,1)$, which at the level of the algebra
is isomorphic to $sl(2,\mathbb{R})\approx su(1,1)$. Let us choose
global coordinates $(t,\theta)$ for $dS_{2}$, \[
ds^{2}=-dt^{2}+\cosh^{2}t~d\theta^{2}.\]
When acting on scalar field modes, the $sl(2,\mathbb{R})$ generators
take the form \begin{eqnarray}
L_{0} & = & \cos\theta\frac{\partial}{\partial t}-\sin\theta\tanh t\frac{\partial}{\partial\theta}\nonumber \\
L_{-1}-L_{1} & = & 2\sin\theta\frac{\partial}{\partial t}+2\cos\theta\tanh t\frac{\partial}{\partial\theta}\nonumber \\
L_{1}+L_{-1} & = & 2\frac{\partial}{\partial\theta}\label{eq:vgens}\end{eqnarray}
 and satisfy the Virasoro algebra\begin{equation}
[L_{m},L_{n}]=(m-n)L_{m+n}~.\label{eq:vira}\end{equation}

We will now show that the representation of $su(1,1)\approx sl(2,\mathbb{R})$
one gets when quantizing a scalar field in de Sitter, in the range
of masses under consideration ($m>1/2$), is in fact a principal series
representation. These representations are part of a larger family
of representations labelled by a continuous complex parameter $\tau$
and an index $\varepsilon$ that can take the discrete values $0$
and $1/2$; we will be interested only in the case $\varepsilon=0$.
They can be realized \cite{Klimyk:1991fg} in terms of generators
$H$, $X_{+}$, $X_{-}$ that act on periodic functions on the circle,
$f(\theta)$ with $\theta\in[0,2\pi)$. Choosing $e^{-ik\theta}$,
$k\in\mathbb{Z}$ as a basis on this space of functions, the action
of the generators is given by\begin{eqnarray}
He^{-ik\theta} & = & 2ke^{-ik\theta},\nonumber \\
X_{+}e^{-ik\theta} & = & (k-\tau)e^{-i(k+1)\theta},\label{eq:hrep}\\
X_{-}e^{-ik\theta} & = & -(k+\tau)e^{-i(k-1)\theta}.\nonumber \end{eqnarray}
In terms of differential operators, the generators are\begin{eqnarray}
H & = & 2i\frac{\partial}{\partial\theta}~,\label{eq:hgens}\\
X_{+} & = & ie^{-i\theta}\frac{\partial}{\partial\theta}-\tau e^{-i\theta},\nonumber \\
X_{-} & = & -ie^{i\theta}\frac{\partial}{\partial\theta}-\tau e^{i\theta}.\nonumber \end{eqnarray}
 The principal series representations correspond to the case $\tau=-1/2+i\rho$,
with $\rho$ real \cite{Klimyk:1991fg}. They are unitary with respect
to the canonical inner product \begin{equation}
\left(f_{1}|f_{2}\right)=\frac{1}{2\pi}\int_{0}^{2\pi}d\theta\overline{f_{1}(e^{i\theta})}f_{2}(e^{i\theta})~.\label{eq:hprod}\end{equation}
The associated notion of conjugation is

\begin{equation}
H^{*}=H,\quad X_{+}^{*}=-X_{-},\quad X_{-}^{*}=-X_{+}.\label{eq:hconj}\end{equation}

Now, in the dS/CFT context, one can define a boundary-to-bulk map
by promoting the modes on the circle to modes on de Sitter,\begin{equation}
e^{-ik\theta}\to y_{k}(t)e^{-ik\theta},\label{eq:bulktob}\end{equation}
where the $y_{k}(t)$ are the normalized Euclidean vacuum modes, constructed
explicitly for arbitrary-dimensional de Sitter in (3.37) of \cite{Bousso:2001mw}.
Using (\ref{eq:hprod}) the boundary propagator is simply $\delta_{k,k'}$,
or equivalently $\sum_{k=-\infty}^{\infty}e^{-ik(\theta'-\theta)}=\delta(\theta'-\theta)$
in coordinate space. The bulk Wightman propagator in the Euclidean
vacuum is then recovered from the boundary propagator $\delta_{k,k'}$
as\begin{eqnarray}
\langle E|\phi(t',\theta')\phi(t,\theta)|E\rangle & = & \sum_{k,k'=-\infty}^{\infty}y_{k'}(t')y_{k}^{*}(t)e^{-ik'\theta'+ik\theta}\delta_{k,k'}\nonumber \\
 & = & G_{E}(t',\theta';t,\theta)\label{eq:evacp}\end{eqnarray}
where $G_{E}$ is the Euclidean vacuum propagator as defined, for
example, in \cite{Bousso:2001mw}.

We can act on the set of modes (\ref{eq:bulktob}) with the generators
(\ref{eq:vgens}), to determine the relation with (\ref{eq:hgens}).
This requires the use of a non-trivial hypergeometric function identity,
noted in formula (161) of \cite{Balasubramanian:2002zh}, with the
result

\begin{eqnarray*}
L_{0} & = & h_{+}\cos\theta+\sin\theta\frac{\partial}{\partial\theta}~,\\
L_{-1}-L_{1} & = & 2h_{+}\sin\theta-2\cos\theta\frac{\partial}{\partial\theta}~,\\
L_{1}+L_{-1} & = & 2\frac{\partial}{\partial\theta}~.\end{eqnarray*}
Comparing with (\ref{eq:hgens}), we see that we indeed have a principal
series representation with $\tau=-h_{+}=-1/2-i\mu$, and\begin{eqnarray}
H & = & i\left(L_{1}+L_{-1}\right),\label{eq:htovir}\\
X_{+} & = & \frac{i}{2}\left(L_{1}-L_{-1}\right)+L_{0},\nonumber \\
X_{-} & = & -\frac{i}{2}\left(L_{1}-L_{-1}\right)+L_{0}.\nonumber \end{eqnarray}
For completeness, let us observe that when $m<1/2$ the quantity $\mu$
is pure imaginary and so the parameter $\tau$ is real. As noted in
\cite{Leblond:2002tf}, in the range $0<m<1/2$ the representation
one finds belongs to the supplementary series ($-1<\tau<0$), whereas
for special values in the range $m<0$ one makes contact with the
discrete series representations. These two additional types of representations
are unitary with respect to a different boundary inner product \cite{Klimyk:1997eb}.

Notice that our boundary basis functions $e^{-ik\theta}$ (or, equivalently,
our bulk basis functions $y_{k}(t)e^{-ik\theta}$) are eigenfunctions
not of $L_{0}$, but of $L_{1}+L_{-1}$. Eigenfunctions of $L_{0}$
with eigenvalue $i\omega$ are of the form

\begin{equation}
F_{\omega}(\theta)=\tan^{i\omega}\frac{\theta}{2}\sin^{\tau}\theta~,\label{eq:definiteweight}\end{equation}
with $\omega$ real. These functions are periodic and so admit a Fourier
series expansion; they are however singular at $\theta=0,\pi$. They
satisfy the following orthogonality relation \cite{Balasubramanian:2002zh}\begin{equation}
\int_{0}^{2\pi}d\theta\overline{F_{\omega}(\theta)}F_{\omega'}(\theta)\propto\delta(\omega-\omega')\label{eq:lzint}\end{equation}
 provided the singularities at $\theta=0,\pi$ are regulated in a
suitable way.

A very interesting point is that we see a single irreducible representation
of the conformal group appear, rather than the two distinct highest-weight
representations that appear in the original version of the correspondence
described above. If we were to try to represent a non-trivial $\alpha$-vacuum
other than the Euclidean vacuum (which as explained in \cite{Goldstein:2003qf}
is unlikely to make sense in the interacting theory coupled to gravity),
we would need to use a linear combination of $y_{k}(t)$ and $\overline{y_{k}(t)}$
in (\ref{eq:bulktob}), and two distinct representations with $\tau=-h_{\pm}$
would appear. These representations are actually equivalent, but the
transformation involves a non-trivial change of basis \cite{Klimyk:1991fg},
which takes the form\[
\overline{\left|k\right\rangle }=\frac{\Gamma(h_{-}-k)}{\Gamma(h_{+}-k)}\left|k\right\rangle \]
for $|k\rangle$ a representation satisfying (\ref{eq:hrep}) with
$\tau=-h_{+}$, and $\overline{|k\rangle}$ a representation satisfying
(\ref{eq:hrep}) with $\tau=-h_{-}$. In coordinate space, this is
rewritten \cite{Balasubramanian:2002zh}\[
\overline{f_{k}}(\theta)\propto\int_{0}^{2\pi}d\theta'\left(1-\cos(\theta-\theta')\right)^{-h_{-}}e^{-ik\theta'}~,\]
 i.e., for the basis $e^{-ik\theta}$ satisfying (\ref{eq:hrep})
with $\tau=-h_{+}$, the basis $\overline{f_{k}}$ will satisfy the
same relations with $h_{+}\to h_{-}$. Note the appearance of the
propagator of a conformal primary of weight $h_{-}$ in this change
of basis.

It is easy to check that the Klein-Gordon inner product for a free
quantum scalar field reduces to the inner product (\ref{eq:hprod}),
up to a positive normalization constant (the simplest way is to look
at a time slice that approaches $\mathcal{I}^{-}$). The adjoint of
the generators with respect to this inner product is therefore given
by (\ref{eq:hconj}), which implies\[
L_{n}^{*}=-L_{n}~.\]
This notion of conjugation is also emphasized in \cite{Balasubramanian:2002zh},
and discussed in \cite{Leblond:2002tf}; it amounts to the statement
that the $SL(2,\mathbb{R})$ group elements, $\exp(c_{n}L_{n})$ with
$c_{n}\in\mathbb{R}$, are unitary. Notice that it differs from the
definition of adjoint considered in \cite{Bousso:2001mw,Witten:2001kn},
$L_{n}^{*}=L_{-n}$. The difference between these two definitions
becomes even more significant in the higher-dimensional case: quantization
of a scalar field in $d$ dimensions yields a unitary principal series
representation of the conformal group in $d-1$ Euclidean dimensions,
$SO(d,1),$ whereas the notion of conjugation employed in \cite{Bousso:2001mw,Witten:2001kn}
makes contact instead with the Lorentzian conformal group $SO(d-1,2)$
\cite{Balasubramanian:2002zh}.

To summarize, we have seen that scalar fields in de Sitter yield principal
series representations of $SU(1,1)\approx SL(2,\mathbb{R})$. All
these representations are irreducible for generic $m$, infinite-dimensional
and without highest or lowest weights. We have also constructed a
bulk-to-boundary map at the level of the field modes: field configurations
on the boundary are mapped into linear combinations of positive-frequency
Euclidean vacuum modes. 

It is natural then to propose a new version of the dS$_{2}$/CFT$_{1}$
correspondence, where bulk correlation functions of a scalar field
are determined by matrix elements in a CFT that lives on a circle,
built out of the unitary principal series representation with $\tau=-h_{+}$.
It is most natural to think of this correspondence in terms of duality
of the bulk theory with a CFT living on a single spatial boundary
in the infinite past $\mathcal{I}^{-}$, where the bulk to boundary
map (\ref{eq:bulktob}) matches the definition of the Euclidean vacuum
(\ref{eq:evacp}). Here we differ with the interpretation proposed
in \cite{Balasubramanian:2002zh} in terms of two entangled CFT's
living on both past and future infinities of de Sitter. Formulating
the correspondence using a single boundary also allows for the possibility
that the CFT  may describe states that are asymptotically de Sitter
in the past, but do not evolve to an asymptotically de Sitter final
state (or vice versa if one takes the boundary at $\mathcal{I}^{+}$).
Aside from some minor differences in the formulas, all we have said
generalizes to the general case dS$_{d}$/CFT$_{d-1}$.

\section{Quantum dS/CFT Proposal \label{sec:qdscft}}

The representations of the $SO(d,1)$ group discussed above are infinite-dimensional,
and so do not naturally explain the finiteness of the horizon entropy
in the dS/CFT framework. To ameliorate this problem, we will show
for the case of $dS_{2}$ that a quantum deformation of the symmetry
group yields finite-dimensional unitary representations that go over
to the unitary principal series representations in the classical limit.

\subsection{Quantum groups\label{sub:Quantum-groups}}

Our basic building block will be the quantum group $SL_{q}(2,\mathbb{C})$,
or more precisely, the universal enveloping algebra $U_{q}(sl(2,\mathbb{C}))$
with complex coefficients built out of generators $K$, $X_{+}$,
$X_{-}$ satisfying \cite{Klimyk:1997eb}

\begin{equation}
KK^{-1}=K^{-1}K=1,\qquad KX_{\pm}K^{-1}=q^{\pm2}X_{_{\pm}},\qquad[X_{+},X_{-}]=\frac{K-K^{-1}}{q-q^{-1}}~.\label{eq:qalg}\end{equation}
 The universal enveloping algebra consists of the space spanned by
the monomials \begin{equation}
(X_{+})^{n}K^{m}(X_{-})^{l}\label{eq:monomial}\end{equation}
 with $m\in\mathbb{Z}$ and $n,l$ non-negative integers. It is a
Hopf algebra with comultiplication $\Delta$ defined as\begin{eqnarray*}
\Delta(K) & = & K\otimes K\\
\Delta(X_{+}) & = & X_{+}\otimes K+1\otimes X_{+}\\
\Delta(X_{-}) & = & X_{-}\otimes1+K^{-1}\otimes X_{-}\end{eqnarray*}
and with antipode $S$ and counit $\varepsilon$ defined as\begin{eqnarray*}
S(K) & = & K^{-1},\qquad S(X_{+})=-X_{+}K^{-1},\qquad S(X_{-})=-KX_{-},\\
\varepsilon(K) & = & 1,\qquad\varepsilon(X_{\pm})=0.\end{eqnarray*}
It is also useful to define a related comultiplication,\[
\Delta'\equiv\sigma\circ\Delta,\qquad\sigma(a\otimes b)\equiv b\otimes a~.\]
Roughly speaking, we can think of the element $K$ as $q^{H}$, where
$H$ becomes the usual Cartan generator of $SL(2)$ when we take the
classical $q\to1$ limit. Written in terms of $H$, the algebra becomes\[
[H,X_{\pm}]=\pm2X_{\pm},\qquad[X_{+},X_{-}]=\frac{q^{H}-q^{-H}}{q-q^{-1}}~.\]
However, it will be important later that the quantum group involves
the restriction to products of generators of the form (\ref{eq:monomial}).
We refer the reader to \cite{Chari:1994pz,Klimyk:1997eb} for an introduction
to quantum groups. 

We will be interested in defining a notion of conjugation on these
generators, in order to specify inner products. This pairing is known
as a {*}-structure. In the mathematics literature, this is usually
defined to preserve the form of the comultiplication, in the sense
that $\Delta(a^{*})=(*\otimes*)\Delta(a)$. We will be interested
however in a generalization of this notion of {*}-structure where
$\Delta(a^{*})=(*\otimes*)\Delta'(a)$, which is discussed extensively
in the introductory sections of \cite{Buffenoir:1997ih}. There exist
a number of different choices of this {*}-structure for $U_{q}(sl(2,\mathbb{C}))$,
which are discussed in \cite{Buffenoir:1997ih,Steinacker:1998kv,Steinacker:1999xu}.
We will be interested in the specific choice\begin{equation}
X_{\pm}^{*}=-X_{\mp},\qquad K^{*}=K^{-1}~,\label{eq:starstruc}\end{equation}
defined for $q$ a root of unity. This is the quantum counterpart
of (\ref{eq:hconj}), which as we saw in Section \ref{sec:newdS/CFT}
is the notion of conjugation relevant to a scalar field on $dS_{2}$.
The map (\ref{eq:starstruc}) is anti-linear (acts by complex conjugation
on c-numbers), involutive and is an anti-morphism (reverses the order
of generators). The Hopf algebra $U_{q}(sl(2,\mathbb{C}))$ combined
with this {*}-structure is known as $U_{q}(su(1,1))$, which again
is to be thought of as an enveloping algebra with complex coefficients.
For the special case that $q$ is a root of unity, we can extract
a real subalgebra of this Hopf algebra, that we denote $U_{q}(su(1,1))_{\mathbb{R}}$.
This is done by defining a map \begin{equation}
\theta(X^{\pm})=-X^{\pm},~\theta(K)=K^{-1},\label{eq:theta}\end{equation}
and showing that the restriction $a^{*}=\theta(a)$ where $a\in U_{q}(su(1,1))$
is compatible with the algebra and comultiplication structure. 

In more detail: one may define the basis of generators\[
J_{3}=iH,~J_{1}=X^{+}+X^{-},\mathrm{~and}~J_{2}=i(X^{+}-X^{-})~,\]
which become a canonical basis for $su(1,1)$ in the classical limit
$q\to1$. The {*}-structure (\ref{eq:starstruc}) defines an involution
of the algebra $U_{q}(sl(2,\mathbb{C}))$. The map $\theta$ (\ref{eq:theta})
maps the Hopf algebra into itself, provided we permute factors in
the comultiplication (i.e., $\theta$ is an anti-comorphism). It is
then easy to check that \[
J_{i}^{*}=-J_{i}\]
for $i=1,2,3$ and that this restriction is compatible with the Hopf
algebra structure. The algebra generated by the $J_{1},J_{2}$ and
$K=q^{-iJ_{3}}$ with real coefficients is then a real Hopf algebra,
which we denote $U_{q}(su(1,1))_{\mathbb{R}}$.

For $q$ a root of unity (for simplicity we will mostly consider the
case $q=e^{2\pi i/N}$ with $N$ odd%
\footnote{The main difference for even $N$ is that the exponents in (\ref{eq:cyclicrel})
, and consequently the dimensions of the irreducible representations
discussed in the next subsection, become $N/2$ \cite{Klimyk:1997eb}.%
}), the center of the algebra involves not just the usual quadratic
Casimir \[
C=X_{-}X_{+}+\frac{Kq+K^{-1}q^{-1}}{(q-q^{-1})^{2}}\]
but also the elements \begin{equation}
X_{+}^{N},~X_{-}^{N},~K^{N},~\mathrm{and}~K^{-N}~.\label{eq:cyclicrel}\end{equation}
This implies that any irreducible representation of the algebra is
finite-dimensional \cite{Klimyk:1997eb}.

\subsection{Classical limit of the quantum group representation}

For $q$ a root of unity (we will take $q=e^{2\pi i/N}$ with $N$
odd) there exists a class of finite-dimensional irreducible representations
of the quantum group that can be realized on the $N$-dimensional
basis $|m\rangle$ with $m=0,\cdots,N-1$ \cite{Klimyk:1997eb} and
parametrized by the complex numbers $a,b,\lambda$:\begin{eqnarray*}
K|m\rangle & = & q^{-2m}\lambda|m\rangle\\
X_{+}|m\rangle & = & \left(ab+\frac{q^{m}-q^{-m}}{q-q^{-1}}\frac{\lambda q^{1-m}-\lambda^{-1}q^{m-1}}{q-q^{-1}}\right)|m-1\rangle\\
X_{-}|m\rangle & = & |m+1\rangle\end{eqnarray*}
supplemented by the additional cyclic operations\begin{equation}
X_{+}|0\rangle=a|N-1\rangle~,\qquad X_{-}|N-1\rangle=b|0\rangle~.\label{eq:cyclicop}\end{equation}
For $a,b\neq0$ there are no highest- or lowest-weight states and
the representation is called cyclic. 

To try to establish a connection with the principal series representations,
let us take the $q\to1$ classical limit of the above expressions,
with $\lambda=q^{2l}$, $K=q^{H}$ and allowing for a change in normalization
of the basis elements, $|m\rangle\to B(m)|m\rangle$:\begin{eqnarray}
H|m\rangle & = & 2(l-m)|m\rangle\nonumber \\
X_{+}|m\rangle & = & \left(ab+m(2l+1-m)\right)\frac{B(m-1)}{B(m)}|m-1\rangle\nonumber \\
X_{-}|m\rangle & = & \frac{B(m+1)}{B(m)}|m+1\rangle\label{eq:hactt}\end{eqnarray}
Setting $B(m+1)/B(m)=m-\tau-l$, \begin{equation}
ab=\tau^{2}+\tau-l^{2}-l\label{eq:abeqn}\end{equation}
 and defining a new basis $|k\rangle'=|l-m\rangle$, (\ref{eq:hactt})
becomes\begin{eqnarray*}
H|k\rangle' & = & 2k|k\rangle'\\
X_{+}|k\rangle' & = & (k-\tau)|k+1\rangle'\\
X_{-}|k\rangle' & = & -(\tau+k)|k-1\rangle'\end{eqnarray*}
 as for the principal series representation (\ref{eq:hrep}). This
makes sense provided we identify $l$ with an integer. Since $\tau$
is complex, $ab$ is in general complex. Furthermore, since $m=0,\cdots,N-1$
we get $k=l-N+1,\cdots,k$. Therefore if we take $l=(N-1)/2$ (odd
$N$) we get $k=-(N-1)/2,\cdots,(N-1)/2$ which gives us the principal
series basis as $N\to\infty$. Notice that up to these conditions
on $l$ and (\ref{eq:abeqn}), the individual values of $a$ and $b$
are undetermined. As we will see in the next subsection, invariance
of the inner product fixes $a$ and \textbf{$b$} up to a phase, and
this phase drops out of the product $ab$ which appears in the action
of the generators (\ref{eq:hactt}). 

In the basis where $L_{1}+L_{-1}=-iH$ is diagonal, we approach the
classical principal series representation in a smooth way. There is
a subtlety however, if we attempt to change basis to diagonalize the
operator $L_{0}=(X_{+}+X_{-})/2$ and then take the $q\to1$ limit.
Because $X_{\pm}^{N}\propto1$, the spectrum of $X_{\pm}$ is $N$
evenly spaced points around a circle centered at the origin. Therefore
the spectrum of $L_{0}$ will also be made up of $N$ discrete points,
and it turns out their spacing remains constant as $N\to\infty$,
so one does not reproduce the continuous spectrum of $L_{0}$ expected
in the classical limit (\ref{eq:definiteweight}). This implies the
$q\to1$, $N\to\infty$ limit does not commute with the regularization
needed to make sense of the completeness relation (\ref{eq:lzint}).

\subsection{Unitarity}

We wish to investigate unitarity of the cyclic quantum group representations
under the conditions (\ref{eq:abeqn}) and $l=(N-1)/2$. For the unitary
principal series representations in the classical limit we have $\tau=-1/2+i\rho$,
which implies $ab$ is always a negative real number. The eigenvalues
of $H$ are real, since $l$ is an integer. It remains to examine\begin{eqnarray}
\left\langle X_{+}m|X_{+}m\right\rangle  & = & -\left\langle m|X_{-}X_{+}|m\right\rangle \nonumber \\
 & = & -\left(ab+\frac{q^{m}-q^{-m}}{q-q^{-1}}\frac{\lambda q^{1-m}-\lambda^{-1}q^{m-1}}{q-q^{-1}}\right)\langle m|m\rangle\label{eq:unorm}\end{eqnarray}
where we have used the notion of conjugation defined by the {*}-structure
(\ref{eq:starstruc}). Substituting in for $ab$ and using $\bar{\tau}=-1-\tau$,
we need to check whether\[
v=l^{2}+l+|\tau|^{2}-\left(\frac{q^{m}-q^{-m}}{q-q^{-1}}\frac{\lambda q^{1-m}-\lambda^{-1}q^{m-1}}{q-q^{-1}}\right)>0.\]
This can be expressed as \[
v=l^{2}+l+|\tau|^{2}-\frac{\sin\left(\frac{2\pi(l-k)}{2l+1}\right)\sin\left(\frac{2\pi(l+1+k)}{2l+1}\right)}{\sin^{2}\left(\frac{2\pi}{2l+1}\right)}=l^{2}+l+|\tau|^{2}+\frac{\sin^{2}\left(\frac{2\pi(l-k)}{2l+1}\right)}{\sin^{2}\left(\frac{2\pi}{2l+1}\right)}\]
which is manifestly positive. The same is true for the special cases
on the edges ($m=0,N-1$). Similar results are obtained for $\left\langle X_{-}m|X_{-}m\right\rangle $.
There is one additional relation that comes from demanding invariance
of the inner product under $X_{+}^{N}$\[
\langle X_{-}^{N}0|0\rangle-\langle0|(X_{-}^{N})^{*}0\rangle=0\]
which leads to the condition\[
|b|^{2}=-\prod_{j=0}^{N-1}s(j)\]
where $s(j)\equiv ab+\frac{q^{m}-q^{-m}}{q-q^{-1}}\frac{\lambda q^{1-m}-\lambda^{-1}q^{m-1}}{q-q^{-1}}$.
This fixes $a$ and $b$ up to an overall phase. Under these conditions,
the cyclic irreducible representations are unitary for arbitrary $N$.

\subsection{qdS/CFT proposal}

Underlying our proposal that, at the quantum level, the isometry group
of two-dimensional de Sitter space should be $q$-deformed, lies of
course the idea that $dS_{2}$ itself should be similarly deformed,
to produce a geometry on which $U_{q}(su(1,1))_{\mathbb{R}}$ has
a natural action. At the classical level, $dS_{2}$ can be understood
as the quotient of the isometry group $SU(1,1)$ by the non-compact
$U(1)$ associated with one of the boost generators, and so, at the
quantum level, one is naturally led to consider $U_{q}(su(1,1))_{\mathbb{R}}/U(1)$.
Similar constructions have been explored in \cite{Jevicki:2000ty,Ho:1999bn}.
We will make some additional comments about this $q$- deformed geometry
in the concluding section; but a more detailed analysis is left for
future work. 

Our overall proposal is then that the $N$-dimensional representations
described above can be used to formulate a new correspondence between
a gravitational theory in a bulk $q$-deformed de Sitter geometry
and a $q$-deformed holographic CFT. We emphasize that the representations
in question are unitary for arbitrary $N$, and in the $N\to\infty$
limit, they make contact with the reformulation of the classical dS/CFT
correspondence in terms of unitary principal series representations
discussed in Section \ref{sec:newdS/CFT}. We believe this provides
a promising microscopic framework to account for the finite entropy
of de Sitter space.

\subsection{Relation to \char`\"{}The trouble with de Sitter space\char`\"{}}

\label{sec:appendixB}

Now that we have a $q$-deformed version of the theory with finite
dimensional representations, we can reanalyze the arguments of \cite{Goheer:2002vf,Goheer:2003tx}
arguing for quasi-periodic matrix elements for any description of
de Sitter compatible with finite entropy, hermiticity of the static
patch Hamiltonian, and covariance under the classical symmetries.
In the static patch with coordinates\[
ds^{2}=\frac{1}{\cosh^{2}r}\left(-dt_{s}^{2}+dr^{2}\right)\]
the $sl(2,\mathbb{R})$ generators take the form\begin{eqnarray*}
L_{0} & = & -\frac{\partial}{\partial t_{s}}\\
L_{-1}-L_{1} & = & -2\cosh t_{s}\sinh r\frac{\partial}{\partial t_{s}}-2\sinh t_{s}\cosh r\frac{\partial}{\partial r}\\
L_{1}+L_{-1} & = & 2\cosh t_{s}\sinh r\frac{\partial}{\partial t_{s}}+2\cosh t_{s}\cosh r\frac{\partial}{\partial r}\end{eqnarray*}
Thus the static patch Hamiltonian is to be identified with \begin{equation}
H_{s}=-iL_{0}=-i(X_{+}+X_{-})/2~.\label{eq:sham}\end{equation}
The argument of \cite{Goheer:2002vf,Goheer:2003tx} proceeds by analyzing
the general matrix element of a hermitian boost generator $L=iL_{-1}=(X_{-}-X_{+}+H)/2$,
which obeys\[
[H_{s},L]=iL.\]
The classical argument proceeds by studying\[
\left\langle \psi|e^{iH_{s}t}e^{iL}e^{-iH_{s}t}|\psi\right\rangle =\left\langle \psi|e^{iLe^{-t}}|\psi\right\rangle \]
with $|\psi\rangle$ a general state. This matrix element approaches
$1$ as $t\to\infty$. On the other hand, under the assumption that
$H_{s}$ has a discrete spectrum, the authors of \cite{Goheer:2002vf,Goheer:2003tx}
show the matrix element must be quasi-periodic in time, and so in
particular cannot approach a constant. 

In our $q$-deformed framework, however, $e^{-iL}$ is not in the
universal enveloping algebra, so the argument does not go through.
Instead one is restricted to operators built out of products of $K$,
$K^{-1}$, $X_{+}$ and $X_{-}$. Since the spectrum of $H_{s}$ in
our representations is manifestly discrete, correlation functions
are guaranteed to display the expected quasi-periodicity. The main
virtue of our approach is that this is achieved without giving up
the hermiticity of $H_{s}$.

\section{Discussion\label{sec:Discussion}}

In this paper we have made two main points. First, we have emphasized
that, as observed in \cite{Balasubramanian:2002zh}, the dS/CFT correspondence
must be formulated in terms of principal series representations of
the isometry/conformal group, as opposed to the standard highest-weight
representations usually considered in CFT. Such a reformulation of
dS/CFT is natural from the bulk point of view, since quantization
of a scalar field on $dS_{d}$ yields representations of the former,
and not the latter, type. In particular, the ordinary Klein-Gordon
inner product directly coincides with the scalar product of the principal
series representations, and differs from the one considered in \cite{Bousso:2001mw,Witten:2001kn},
which is on the other hand associated with the usual CFT notion of
adjoint. But the reformulation is in fact also natural from the perspective
of the boundary theory, because the putative dual CFT lives on a $(d-1)$-dimensional
space that is Euclidean from the start, and is not as in the usual
case obtained by analytic continuation from an originally Lorentzian
spacetime. The relevant conformal group is consequently $SO(d,1)$
and not $SO(d-1,2)$ \cite{Balasubramanian:2002zh}. Most importantly,
the principal series representations are unitary, so in the new formulation
one avoids the problems associated with the non-unitarity of the highest-weight
representations that appear in \cite{Strominger:2001pn}.

Of course, one of the motivations of \cite{Witten:2001kn,Bousso:2001mw}
for concentrating on a bulk inner product that differs from the ordinary
one was to try to obtain a framework that departs from the standard
perturbative quantization of the scalar field on $dS$, and has consequently
at least some chance of making contact with the finite-dimensional
(or at least discrete-energy) Hilbert space that the finite entropy
of de Sitter seems to hint at \cite{Bousso:2000nf,Banks:2002wr,Banks:2003ta,Goheer:2002vf,Dyson:2002nt}.
Our second main point in this paper has been that it is possible to
achieve this goal without losing contact with the principal series
story, as long as we are willing to give up the classical symmetries
and trade them for a $q$-deformed version, with $q$ a root of unity.
We gladly pay this price because in exchange we have obtained a finite-dimensional
framework that is \emph{manifestly unitary}.

Thus far, a precise description of the $q$-deformed de Sitter geometry
and its spacetime physics is lacking; but let us make a few general
comments based on the structure of the algebra and the representations
we are considering. We hope to return to this set of issues in more
detail in future work. The cyclic relation (\ref{eq:cyclicrel}) implies
that all irreducible representations are finite-dimensional. The index
$k$ that labels our basis of states $|k\rangle'$ is interpreted
as momentum around the circle in the classical limit, so the $q$-deformation
can be thought of as enforcing an ultraviolet cutoff on this momentum.
Thus the Euclidean boundary space of the dS/CFT correspondence, dual
to this momentum, can be roughly thought of as a discrete set of points%
\footnote{See \cite{Pouliot:2003vt} for an interesting set of related examples
of $q$-deformed spaces.%
}. Likewise, one deduces the spectrum of the operators $X^{+}$ and
$X^{-}$ must also be discrete with an ultraviolet cutoff of order
$N$. This implies the same is true of the Hamiltonian (\ref{eq:sham}),
so it seems the time direction of the bulk de Sitter spacetime also
becomes discretized. 

Systems with finite dimensional Hilbert spaces undergo Poincar\'e
recurrence, as has been discussed extensively in the context of de
Sitter spacetime in \cite{Banks:2002wr,Goheer:2002vf,Dyson:2002nt,Dyson:2002pf,Goheer:2003tx}.
If the dimension of the Hilbert space is $e^{S}$ where $S$ is the
statistical mechanical entropy, then there will be a recurrence time
$t_{r}\propto e^{S}$. Likewise energy measurements will always be
uncertain at order $1/t_{r}=e^{-S}$. In our setup, if the dimension
of Hilbert space is of the same order of that of a single $q$-deformed
representation, we would obtain the same relations with $N=e^{S}$.

Of course so far we have only considered the properties of a single
irreducible representation. One might expect the full Hilbert space
to be built out of arbitrary tensor products of these representations,
which would enlarge the number of states to infinity. Interactions
are built using the fusion rules for these representations, which
have been thoroughly studied in \cite{Arnaudon:1994ig,Keller:1990tg}.
However as we have said, space should be thought of as a finite number
of points, so we are far from the situation where we have a spacetime
with a well-defined asymptotic region where multi-particle states
can be built neglecting interactions. Nevertheless, even if we assume
this infinite-dimensional Hilbert space is the correct description,
the energy eigenvalues can remain discrete, which can still lead to
Poincar\'e recurrence \cite{Dyson:2002pf}, and, more importantly,
can be compatible with the finite entropy of de Sitter \cite{Goheer:2002vf}.
In the static patch, a cutoff is naturally implemented in the form
of a Boltzmann weighting of states at finite temperature, so the entropy
can be accounted for by a finite number of states below an energy
of order the Hubble scale \cite{Goheer:2002vf} %
\footnote{We should also bear in mind that the condition that the spacetime
be close to de Sitter in the past and future in a suitable classical
limit could translate into a cutoff on the number of states relevant
to the de Sitter entropy calculation. Conversely, the full Hilbert
space may then contain states that do not approach de Sitter in the
past and/or future as one takes the classical limit.%
}.

Let us describe this in a little more detail. To isolate the states
defined on the causal diamond of an observer on the north pole of
de Sitter, it is necessary to trace over modes on the southern diamond
\cite{Bousso:2001mw}. This can be implemented by viewing the modes
on the southern diamond as a thermofield double of modes on the northern
diamond \cite{Goheer:2002vf}. Provided we start in the Euclidean
vacuum, integrating out the southern modes then gives rise to a thermal
Boltzmann density matrix for the northern diamond modes. To make contact
with this paper, eigenfunctions of $L_{0}$ with $\omega>0$ should
be viewed as static patch modes on the northern diamond, and likewise
$\omega<0$ modes correspond to the southern diamond. The bulk to
boundary map places these modes in the Euclidean vacuum, so for the
classical version of the dS/CFT correspondence described in section
\ref{sec:newdS/CFT}, the same story will carry over. With a better
understanding of the $q$-deformed de Sitter geometry, and the bulk
to boundary map in particular, we hope a similar story will also hold
in the $q$-dS/CFT case.

Most of the ideas studied in this paper generalize directly to higher-dimensional
de Sitter space. In particular, the classical de Sitter isometry groups
$SO(d,1)$ have unitary principal series representations. The unitary
norm is naturally defined on functions on the $d-1$ sphere, which
we identify with the holographic boundary. The adjoint on this norm
again differs from that proposed in \cite{Witten:2001kn,Bousso:2001mw},
as also emphasized in \cite{Balasubramanian:2002zh}. We conjecture
there will likewise be a sensible $q$-deformation of the classical
isometry group and that the associated $q$-deformed holographic conformal
field theory based on a deformation of the unitary principal series
representations will be dual to a gravitational theory in the bulk
$q$-deformed de Sitter geometry. We hope to further elaborate on
the details of this higher-dimensional correspondence in future work.

\begin{acknowledgments}
D.L. thanks Hugo Compe\'an, Alex Maloney, Don Marolf, Djordje Minic,
Philippe Pouliot, Sanjaye Ramgoolam and Andrew Strominger for helpful
discussions, and the Departamento de F\'{\i}sica, CINVESTAV and KITP
for hospitality. A.G. is grateful to Chryssomalis Chryssomalakos,
Anatoli Klimyk and David Vergara for useful conversations. D.L. is
supported in part by DOE grant DE-FE0291ER40688-Task A, NSF U.S.-Mexico
Cooperative Research grant 0334379, and NSF grant PHY99-07949. The
work of A.G. is additionally supported by Mexico's National Council
of Science and Technology (CONACyT).
\end{acknowledgments}
\bibliographystyle{apsrev}
\clearpage\addcontentsline{toc}{chapter}{\bibname}\bibliography{qdscft}

\end{document}